\documentclass[letterpaper,10pt]{article}
\usepackage{osameet2}

\usepackage{amsmath,amssymb}

\usepackage[dvips,colorlinks=true,bookmarks=false,citecolor=blue,urlcolor=blue]{hyperref} 
\usepackage{subfigure}

\begin{document}

\title{Robust Phase Estimation of Squeezed State}

\author{Shibdas Roy*, Ian R. Petersen, Elanor H. Huntington}
\address{School of Engineering and Information Technology, University of New South Wales, Canberra, Australia.}
\email{*shibdas.roy@student.unsw.edu.au}

\begin{abstract}
Optimal phase estimation of a phase-squeezed quantum state of light has been recently shown to beat the coherent-state limit. Here, the estimation is made robust to uncertainties in underlying parameters using a robust fixed-interval smoother.
\end{abstract}

\vspace*{2mm}

\section{Introduction}
Precise estimation and tracking of a randomly varying optical phase is key to applications like communication \cite{CHD}, and metrology \cite{GLM1}. Optimal smoothed estimation of a widely varying phase under the influence of a stochastic Ornstein-Uhlenbeck (OU) process for a phase-squeezed beam noticeably surpasses the maximum accuracy that can be obtained for coherent light by $15 \pm 4\%$ \cite{YNW}. But, the estimation is very sensitive to fluctuations in the parameters underlying the phase noise or squeezing, and is desired to be made robust to such uncertainties. We have already demonstrated in Ref. \cite{RPH2}, superior phase estimation as compared to the optimal case for a coherent state \cite{TW}, using a robust fixed-interval smoother where the uncertain system satisfies a certain integral quadratic constraint (IQC) \cite{MSP}. Here, the robust smoothing technique is applied to the case of squeezed state in comparison to the optimal case considered in Ref. \cite{YNW}.

\section{Model of Adaptive Phase Estimation in Ref. \cite{YNW}}
\subsection{Process Model}
The phase $\phi(t)$ of the continuous optical phase-squeezed beam is modulated with an OU noise process, such that
\vspace*{-2mm}
\begin{equation}
\dot{\phi}(t) = -\lambda\phi(t) + \sqrt{\kappa}\nu(t),
\end{equation}
where $\lambda^{-1}$ is the correlation time of $\phi(t)$, $\kappa$ determines the magnitude of the phase variation and $\nu(t)$ is a zero-mean white Gaussian noise with unity amplitude.

\subsection{Measurement Model}
The phase-modulated beam is measured by homodyne detection using a local oscillator, the phase of which is adapted with the filtered estimate $\phi_f(t)$, thereby yielding a normalized homodyne output current $I(t)$,
\vspace*{-2mm}
\begin{align}
I(t)dt \simeq 2|\alpha|[\phi(t)-\phi_f(t)]dt + \sqrt{\bar{R}_{sq}}dW(t), 
\quad \bar{R}_{sq} = \sigma_f^2 e^{2r_p} + (1-\sigma_f^2)e^{-2r_m},
\end{align}
where $|\alpha|$ is the amplitude of the input phase-squeezed beam, and $dW(t)$ is Wiener noise arising from squeezed vacuum fluctuations. The parameter $\bar{R}_{sq}$ is determined by the degree of squeezing ($r_m \geq 0$) and anti-squeezing ($r_p \geq r_m$) and by $\sigma_f^2 = \langle[\phi(t)-\phi_f(t)]^2\rangle$. We use the measurement appropriately scaled as our measurement model,
\vspace*{-2mm}
\begin{equation}
\theta(t) := 1/\sqrt{\bar{R}_{sq}}[I(t)+2|\alpha|\phi_f(t)] = 2|\alpha| /\sqrt{\bar{R}_{sq}}\phi(t) + \omega(t),
\end{equation}
\vspace*{-2mm}
where $\omega(t) = \frac{dW(t)}{dt}$ is another zero-mean white Gaussian noise with unity amplitude.

\section{Robust Model}

\subsection{Uncertain System}
We employ the robust fixed-interval smoothing technique from Ref. \cite{MSP} as used in Ref. \cite{RPH2}. We introduce uncertainties $\delta_1$ and $\delta_2$, respectively, in the parameters $\lambda$ and $2|\alpha|/\sqrt{\bar{R}_{sq}}$, such that Eq. (2.5) in Ref. \cite{MSP} takes the form:
\vspace*{-3mm}
\begin{align}
\textsf{Process:} \quad \dot{\phi} = (-\lambda + \sqrt{\kappa}\Delta_1 K)\phi + \sqrt{\kappa}\nu, \qquad
\textsf{Measurement:} \quad \theta = \left(2|\alpha|/\sqrt{\bar{R}_{sq}} + \Delta_2 K\right)\phi + \omega,
\end{align}
where $\Delta_1 = \left[\begin{array}{cc}
\delta_1 & 0
\end{array}\right],
\Delta_2 = \left[\begin{array}{cc}
0 & \delta_2
\end{array}\right],|\delta_1| \leq 1, |\delta_2| \leq 1,
K = \left[\begin{array}{c}
-\mu\lambda/\sqrt{\kappa}\\
2\mu|\alpha|/\sqrt{\bar{R}_{sq}}
\end{array}\right]$. $0\leq\mu<1$ is the level of uncertainty.
$\Delta_1$ and $\Delta_2$ satisfy:
$||[\begin{array}{cc}
\Delta_1' Q^{1 \over 2} & \Delta_2' R^{1 \over 2}
\end{array}]|| \leq 1$, such that $Q=R=1$.
The IQC of Eq. (2.4) in Ref. \cite{MSP} for our model is:
\vspace*{-2mm}
\begin{equation}
\int_0^T (w^2 + v^2)dt \leq 1 + \int_0^T ||z||^2 dt,
\end{equation}
\vspace*{-2mm}
where $z = K\phi$ is the uncertainty output, and $w = \Delta_1 K\phi + \nu$ and $v = \Delta_2 K\phi + \omega$ are the uncertainty inputs.

\subsection{Robust vs. Optimal Smoothers for the Uncertain System}
\begin{figure}[h]
\centering
\hspace*{-6mm}
\subfigure[]{\label{fig:subfig:a}
\includegraphics[height=2.2in,width=3.3in]{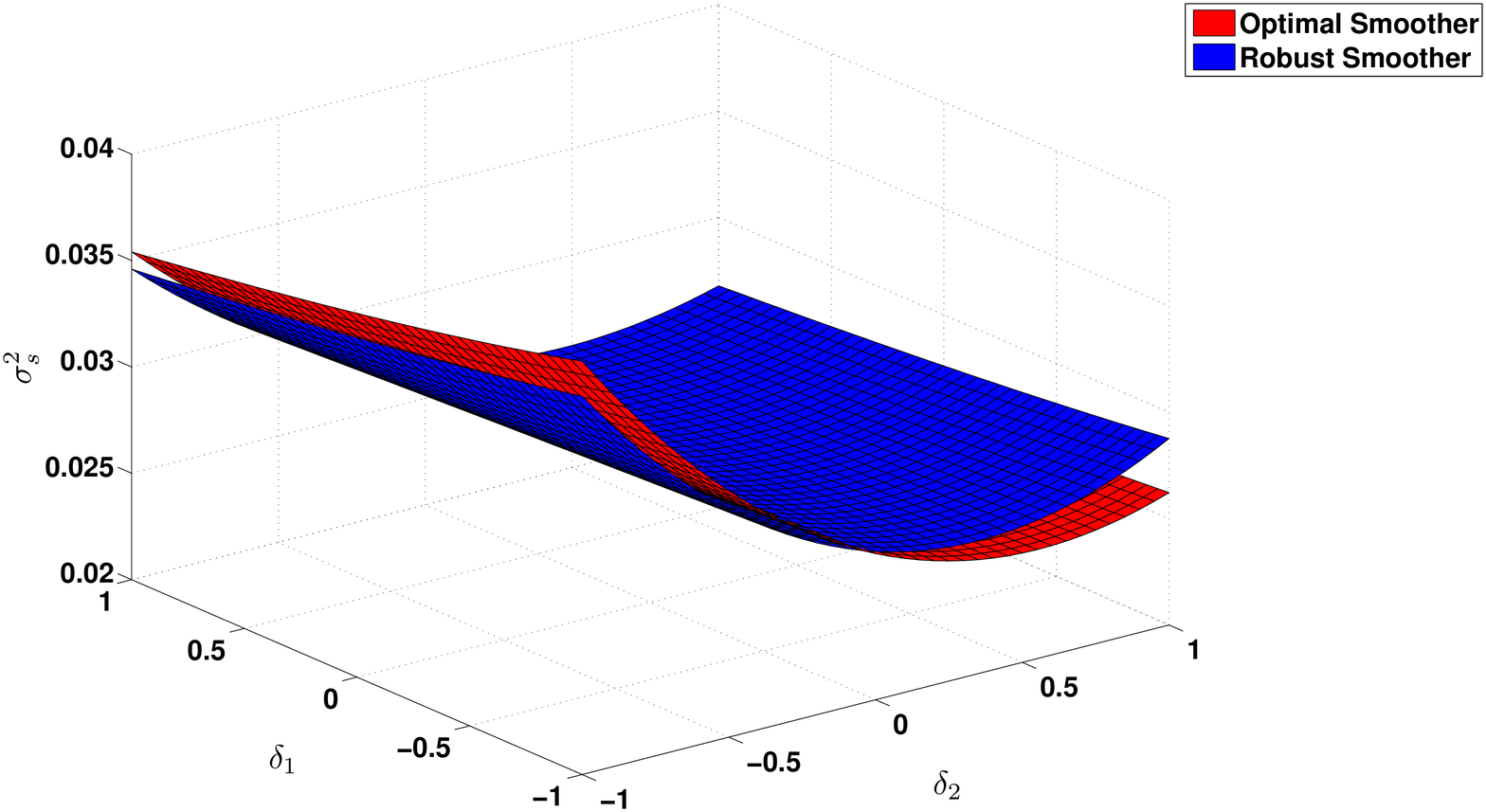}}
\hspace*{-6mm}
\subfigure[]{ \label{fig:subfig:b}
\includegraphics[height=2.2in,width=3.3in]{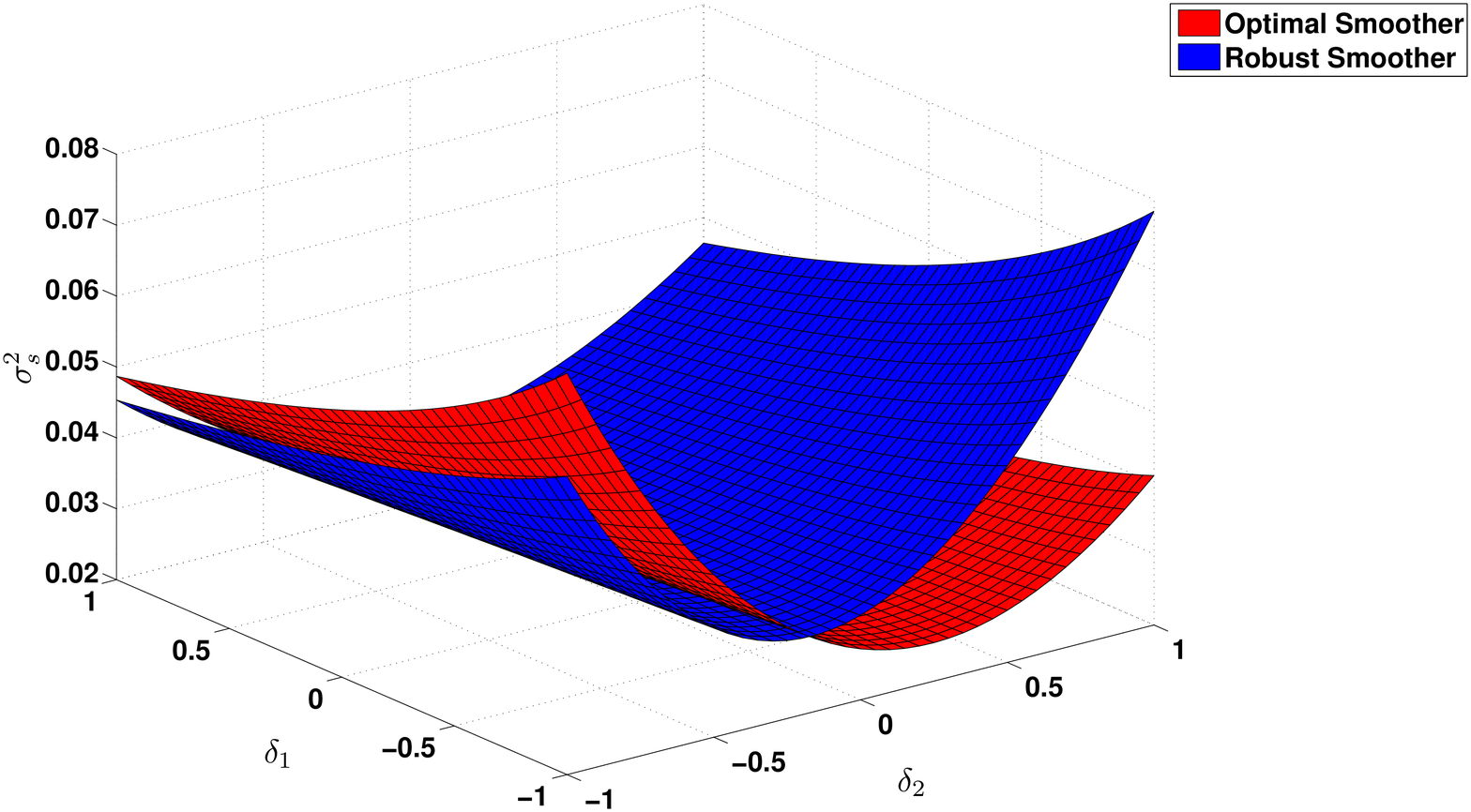}}
\hspace*{-6mm}
\caption{Comparison of Smoothed Errors for Uncertain System: (a) $\mu = 0.2$, (b) $\mu = 0.4$.}
\label{fig:comparison_graphs}
\end{figure}

We use the method from Ref. \cite{RPH2} to compute and compare the errors of the robust and optimal smoothers as a function of $\delta_1$ and $\delta_2$, for various values of $\mu$ and with $\kappa = 1.9 \times 10^4$ rad/s, $\lambda = 5.9 \times 10^4$ rad/s, $|\alpha|^2 = 1.00 \times 10^6 s^{-1}$, $r_m = 0.36$, and $r_p = 0.59$ as used in Ref. \cite{YNW}. Due to the implicit dependence of $\bar{R}_{sq}$ and $\sigma_f^2$, we compute the error $\sigma_s^2$ upon running several iterations until $\sigma_f^2$ matches upto the $6^{th}$ decimal place to that of the prior iteration in each case. Fig. \ref{fig:comparison_graphs} shows the comparison for $20\%$ and $40\%$ uncertainties. Clearly, the robust smoother performs much better than the optimal smoother as $\delta_1$ and/or $\delta_2$ approach $-1$ for all levels of uncertainty.
\vspace*{-2mm}
\vspace*{-2mm}

\section{Conclusion}
This work extends the treatment of robust smoothing to the case of squeezed state. For $\bar{R}_{sq}=1$, it boils down to the coherent state case of Ref. \cite{RPH2} when $\Delta_2 = 0$. These results may be demonstrated experimentally as part of further work.

\bibliographystyle{osajnl}
\bibliography{bibliography}

\end{document}